\documentclass[prl,twocolumn]{revtex4}

\begin{document}

\noindent
{\bf Comment on ``Polynomial-Time
Simulation of Pairing Models on a Quantum Computer''}

\medskip

In a recent Letter\cite{Wu}, Wu {\it et al.} proposed a polynomial-time algorithm (PTA) for
simulations of a wide class of pairing hamiltonians on a NMR quantum computer. While this is an
interesting result we want to recall in this Comment the existence of classical algorithms for
attacking this problem to {\it any} required precision. There are two main sort of algorithms. If
the pairing model (PM) turns out to be integrable then it can be solved \`a la Bethe. This happens
in the reduced BCS model characterized by a single coupling constant, but there are more general
cases reviewed below. The other algorithm is the density matrix renormalization group (DMRG), which
is applicable to both integrable and nonintegrable models.

The basic idea of the Letter is that within the next generation of quantum computers (QC) with up to
10-100 qubits, it will be possible to solve hard quantum many body problems in Hilbert spaces with
$N\sim 10^{2}$ single particle levels and $M\sim 10^{2}$ number of particles. However the authors
seem to have overlooked the aforementioned facts concerning the power of analytical and numerical
methods as the DMRG. While the analytical methods are only applicable to integrable models, the DMRG
is of general applicability.

As mentioned in \cite{Wu}  the PM is of fundamental importance for nuclear structure studies in
medium and heavy nuclei and it was introduced in nuclear physics \cite{Bhor} soon after the work of
Bardeen, Cooper and Schieffer.
Even though the BCS theory explains qualitatively the superconducting phenomenon in finite nuclei,
it suffers from the strong number fluctuations.

More recently the (PM) Hamiltonian has been extensively used to describe the physics of ultrasmall
superconducting grains. It was shown in this context that only by resorting to the exact numerical
solution \cite{Duke1}, using the DMRG, was it possible to describe the disappearance of
superconducting correlations as a function of the grain size. The ground state energy for systems as
large as $N\sim 400$ with Hilbert space dimensions of $\sim 10^{130}$ were treated with high
numerical accuracy (5 significant figures). In particular, the case of $N=100$ was selected to study
the convergence properties of the method showing that the exact ground state energy can be obtained
to seven significant figures with a moderate computational effort (see Table I of Ref.
\cite{Duke1}). The DMRG is a highly accurate iterative method based on the Wilson renormalization
group. In the infinite algorithm procedure used in \cite{Duke1} the number of iterations needed to
obtain the low lying states of a pairing Hamiltonian is $N/2$ which has to be compared with the
$\sim N^4$ computational steps required by PTA of Ref. \cite{Wu}. At each iteration de DMRG
procedure stores the matrix elements of all relevant operators in an optimum subspace of $m$
selected states. Due to the exponential convergence of the method $m$ is tuned to obtain the desired
precision. For a general pairing hamiltonian there are three operators per level, therefore the
total memory usage is $O(3 m^2 N)$ which for reasonable values of $m\sim10^2$ allows the treatment
of systems with $N\sim10^3$. In \cite{Duke1} we used a constant pairing interaction but the DMRG
approach can easily accommodate arbitrary pairing matrix elements, and it is specially suited to
give the ground state and the low lying excited states.

Immediately after this work appeared, it was realized that the PM has been solved exactly by
Richardson in the 1960's \cite{Richa}. Recently \cite {apli}, the exact solution of the PM has been
generalize to a large class of pairing Hamiltonians. The most general pairing Hamiltonian can be
written as
\begin {equation}
H_{p}=\sum_{i}\varepsilon _{i}n_{i}+\sum_{i\neq
j}V_{ij}^{1}~c_{i+}^{\dagger }c_{i-}^{\dagger
}c_{j-}c_{j+}+\sum_{i\neq j}V_{ij}^{2}~n_{i}n_{j} \label{HP}
\end {equation}
where $n_{i}=c_{i+}^{\dagger }c_{i+}+c_{i-}^{\dagger }c_{i-}$ and $i$ runs over the $N$ double
degenerate levels. Note that the third term corresponds to a monopole interaction absent in the
Hamiltonian treated in the Letter. As shown in ref. \cite{apli} there is a wide subset of exactly
solvable Hamiltonians parametrizing the interaction as $\varepsilon _{i}=\epsilon
_{i}-g\sum_{k\left( \neq i\right) }\gamma \left( \epsilon _{i}-\epsilon _{j}\right) \cot \gamma
\left( \eta _{i}-\eta _{j}\right)$, $ V_{ij}^{1}=2g\frac{\gamma \left( \epsilon _{i}-\epsilon
_{j}\right) }{\sin \gamma \left( \eta _{i}-\eta _{j}\right) } $ and $ V_{ij}^{2}=\frac{g}{2}\gamma
\left( \epsilon _{i}-\epsilon _{j}\right) \cot \gamma \left( \eta _{i}-\eta _{j}\right) $. Where
$g$, $\epsilon_i$ and $\eta_i$ are a set of $2 N+1$ free parameters and $\gamma$ can take the values
$0$, $1$ and $-i$ defining the three families of exactly solvable models, the rational, the
trigonometric, and the hyperbolic models respectively. As we see, among the $2N^2-N$ set of free
parameters of the most general and non-integrable Hamiltonian (\ref{HP}), a restriction to a subset
of $6 N +3$ free parameters leads to an exactly solvable Hamiltonian. We believe that most of the
physical problems can be modelled with a pairing Hamiltonian within the integrable subset, but if
some particular case requires a non-integrable pairing Hamiltonian, it can then be solved
numerically to the desired precision using the DMRG.

\medskip

\noindent
J. Dukelsky$^1$, J.M. Rom\'an$^2$ and G. Sierra$^2$

\medskip

$^1$ Instituto de Estructura de la Materia, CSIC, Spain.

$^2$ Instituto de F\'{\i}sica Te\'orica, CSIC-UAM, Spain.

\medskip



\begin{thebibliography}{9}

\bibitem{Wu}  L. A. Wu, M. S. Byrd and D. A. Lidar, Phys. Rev. Lett. {\bf 89}, 057904 (2002).

\bibitem{Bhor} A. Bhor, B. Mottelson and D. Pines, Phys. Rev. {\bf 110}, 936 (1958).

\bibitem{Duke1} J. Dukelsky and G. Sierra, Phys. Rev. Lett. 83, 172 (1999)

\bibitem{Richa}  R.W. Richardson and N. Sherman, Nucl. Phys. {\bf 52}, 221
(1964).

\bibitem{apli} J. Dukelsky, C. Esebbag and P. Schuck, Phys. Rev.
Lett. {\bf 87}, 066403 (2001).


\end{thebibliography}
\end{document}